\documentclass[12pt,a4paper]{article} 

\usepackage[utf8]{inputenc}
\usepackage[T1]{fontenc}
\usepackage{lmodern} 
\usepackage{geometry} 
\usepackage{graphicx}
\usepackage{multirow}
\usepackage{amsmath,amssymb,amsfonts}
\usepackage{amsthm}
\usepackage{mathrsfs}
\usepackage[title]{appendix}
\usepackage{xcolor}
\usepackage{textcomp}
\usepackage{manyfoot}
\usepackage{booktabs}
\usepackage{algorithm}
\usepackage{algorithmicx}
\usepackage{algpseudocode}
\usepackage{listings}
\usepackage{subcaption,caption}
\usepackage{epstopdf}
\usepackage{url}
\usepackage{rotating}
\usepackage{placeins}
\usepackage{float}
\usepackage{enumerate}
\usepackage{setspace}
\usepackage{tikz}
\usepackage{cite}
\usepackage[percent]{overpic}
\usepackage[colorlinks=true, linkcolor=blue, citecolor=red, urlcolor=magenta]{hyperref}

\newtheoremstyle{thmstyleone}%
  {3pt}{3pt}
  {\itshape}
  {}
  {\bfseries}
  {.}
  { }
  {}

\newtheoremstyle{thmstyletwo}%
  {3pt}{3pt}%
  {\normalfont}%
  {}%
  {\bfseries}%
  {.}%
  { }%
  {}%

\newtheoremstyle{thmstylethree}%
  {3pt}{3pt}%
  {\normalfont}%
  {}%
  {\bfseries}%
  {.}%
  { }%
  {\itshape}%

\theoremstyle{thmstyleone}

\theoremstyle{thmstyletwo}

\theoremstyle{thmstylethree}

\begin{document}

\title{Fractal Dimension in Nonlinear Wave Dynamics Governed by a Nonlinear Partial Differential Equation}

\author{
Saugata Dutta\textsuperscript{1}\thanks{Email: \href{mailto:saugatadutta.apd@gmail.com}{saugatadutta.apd@gmail.com}},
Kajal Kumar Mondal\textsuperscript{2}\thanks{Email: \href{mailto:kkmondol@gmail.com}{kkmondol@gmail.com}},
Prasanta Chatterjee\textsuperscript{1}\thanks{Email: \href{mailto:prasantacvb@gmail.com}{prasantacvb@gmail.com}} \\[4pt]
\textsuperscript{1}\small Department of Mathematics, Siksha Bhavana, Visva-Bharati, Santiniketan-731235, India \\
\textsuperscript{2}\small Department of Mathematics, Cooch Behar Panchanan Barma University, Coochbehar-736101, India
}
\date{}
\maketitle
\begin{abstract}
    This work presents a detailed analytical and geometrical investigation of the (2+1)-dimensional Boiti–Leon–Pempinelli system, a nonlinear dispersive model arising in the context of fluid and plasma dynamics. By employing a projective Riccati-based ansatz, a new class of exact solutions is systematically derived. These solutions, when visualized, exhibit intricate geometrical features that evolve across multiple spatial scales. To quantify this complexity, a voxel-based box-counting dimension analysis is conducted on the corresponding surface profiles. The analysis reveals non-integer fractal dimensions that vary with magnification, confirming the self-affine nature of the patterns and highlighting the multiscale structure inherent in the system. Such fractal character is not only of theoretical interest but also reflects real-world behaviors in turbulent plasma flows and fine-scale fluid instabilities. The study thus bridges exact analytical solutions with computational fractal geometry, providing a deeper understanding of the BLP system and its relevance in describing natural phenomena characterized by spatial complexity and multiscale interactions.
\end{abstract}

   \smallskip
\noindent\textit{\textbf{Keywords:}} Boiti–Leon–Pempinelli system; Riccati method; Exact analytical solutions; Fractal structures; Geometric complexity;  Self-similarity; Box-counting dimension



\maketitle

\section{Introduction:}

Nonlinear partial differential equations (NLPDEs) serve as foundational models for describing complex behaviors in physics, biology, chemistry, and other natural sciences. The nonlinearity intrinsic to many such systems presents significant challenges in obtaining exact solutions, which has driven extensive research over recent decades. In this study, a projective equation framework combined with a specific ansatz is employed to derive novel exact solutions of the (2+1)-dimensional Boiti–Leon–Pempinelli (BLP) system. The complex features of these solutions are also investigated within the context of a nonlinear system. The governing BLP system is expressed as:
\begin{equation}\label{1}
\left.
	\begin{array}{cc}
		&u_{ty} = (u^2 - u_x)_{xy} + 2 u_{xx},\\
        &v_t = u_x + 2u v_x.
	\end{array}
    \right \}
\end{equation}
Here, \(u(x,y,t)\) denotes the primary wave field representing the nonlinear dispersive component, while \(v(x,y,t)\) is an auxiliary field coupled to \(u\). This system was originally introduced by Boiti, Leon, and Pempinelli in 1987~\cite{boiti1987spectral}, which generalizes the Korteweg–de Vries (KdV) equation into (2+1) dimensions, incorporating arbitrary spatial and temporal functions. In the limit of reduced dimensionality, the system aligns with the classical dispersive long wave equation, demonstrating its fundamental importance in the theory of nonlinear wave propagation. The system's mathematical structure has made it a central object of study. Garagash~\cite{garagash1994modification} analyzed its Hamiltonian form and identified specific soliton solutions. Later, Lü and Zhang~\cite{lu2004soliton} applied the extended tanh method to derive closed-form solutions. Fang et al.~\cite{fang2005new} expanded this line of research by employing an extended mapping method to reveal fractal and localized patterns in the BLP system. Other significant contributions include Wazwaz and Mehanna~\cite{wazwaz2010comparison}, who investigated traveling wave solutions using exponential and tanh-coth functions, and Aghdaei and Heris~\cite{aghdaei2011exact}, who applied the generalized \((G'/G)\)-expansion technique. Baskonus and Bulut~\cite{baskonus2016exponential} proposed an improved Bernoulli sub-equation method to construct novel solitary waves. Additional developments include the works of El-Shiekh~\cite{elshiekh2017periodic}, who studied solutions under variable coefficients, and Yokus et al.~\cite{yokus2020two}, who utilized the \((1/G')\)-expansion method for hyperbolic-type structures. The analytical exploration of the BLP system has been broadened through numerous methodologies encompassing integrability analysis, Lie symmetries, and various functional transformations~\cite{kumar2015some, ren2007new, silambarasan2021longitudinal, rezazadeh2019new, sulaiman2020three, jhangeer2021new, gao2020complex, pervaiz2020haar, duran2020solitary, subasi2014stability, ahmad2021analytic}.

Despite this progress, limited attention has been paid to the emergence of fractal and self-similar structures within BLP solutions. This study addresses that gap by analyzing the fractal features of the system using voxel-based box-counting methods to estimate their fractal dimension. Such analysis provides deeper insight into the multi-scale and self-repeating structures intrinsic to nonlinear dispersive systems. Among analytical tools, the Riccati method~\cite{li2003nonlinear,chen2003generalized} is particularly effective for generating self-similar, fractal, and localized wave solutions. Chen and Zhang~\cite{chen2003generalized} employed this method to derive soliton-like and non-traveling solutions of the (2+1)-dimensional Boussinesq equation. Ma and Zheng~\cite{ma2006two} adapted Riccati-based mappings to explore variable separation solutions with arbitrary functions in the (2+1)-dimensional dispersive long wave equation, unveiling both conventional and fractal excitations. Ye and Zheng~\cite{ye2007chaotic} extended these ideas to the modified dispersive water wave system, identifying chaotic and fractal waveforms under specific conditions. Z. Li~\cite{li2014new} used a projective Riccati framework to analyze the Broer–Kaup equations and revealed oscillatory and cross-shaped fractal structures. Within the BLP context, the Riccati equation typically assumes the form \begin{equation}\label{2}
    \zeta' = \delta \zeta + \zeta^2,
\end{equation} where \( \zeta' = \frac{d\zeta(q)}{dq} \) and \( q = q(x,y,t) \). This transformation simplifies complex NPDEs into solvable algebraic forms~\cite{conte1992link,chen2004general,kopccasiz2024unveiling}, allowing fractal and self-similar properties to emerge naturally. Reformulation in this manner highlights scale invariance and recursive features inherent to the system.

A defining feature of fractals is their non-integer dimension~\cite{mandelbrot1983fractal,mandelbrot1989fractal,falconer2013fractals}, a measure capturing the geometric intricacy and space-filling capacity of self-similar patterns. In nonlinear wave systems, the fractal dimension quantifies the degree of irregularity and spatial complexity, offering insights into the interplay between nonlinearity and dispersion. Guo et al.~\cite{guo2025fractal} recently explored such fractal solutions in dispersive wave equations, proposing formal methods to compute these dimensions and laying groundwork for future fractal analyses in PDEs.

Motivated by these developments, the present work applies the Riccati-based approach to analyze fractal and self-similar behaviors in the BLP system. Our study focuses particularly on the geometric complexity of resulting solutions through dimension analysis—an aspect not previously addressed. This effort provides new insights into the rich structures emerging in nonlinear PDEs and suggests potential implications for phenomena such as fluid flow, turbulence, and biological transport.

This paper is organized as follows: Section~\ref{S2} presents exact analytical solutions using the Riccati-based method. Section~\ref{S3} discusses the formation of fractal patterns. In Section~\ref{S4}, the fractal dimension of these structures is analyzed. Section~\ref{S5} concludes the study with a summary of the key findings and their broader significance.

\section{Construction of Exact Solutions for the BLP System:}\label{S2}

To derive exact solutions of the BLP system, the projective equation method was employed in combination with a suitable ansatz. This method considers a general nonlinear partial differential equation of the form:
\begin{equation}
P(u, u_t, u_x, u_{xx}, \ldots) = 0,
\end{equation}
where $u = u(x_0, x_1, x_2, \ldots, x_m)$ is treated as the dependent variable, with $x_0 = t$ representing the temporal component and the remaining $x_i$ denoting spatial coordinates. The function $P$ is taken to be a polynomial in $u$ and its partial derivatives.

A solution was proposed in the following generalized form:
\begin{equation}\label{4}
u = A(x,y,t) + \sum_{i=1}^n B_i(x,y,t) \zeta^i(q(x,y,t)) + C_i(x,y,t) \zeta^{i-1}(q(x,y,t)) \sqrt{\delta \zeta + \zeta^2(q(x,y,t))},
\end{equation}
where $\zeta$ is assumed to satisfy a Riccati-type differential equation (see Eq.~\eqref{2}).

By selecting particular forms of the Riccati solution $\zeta$, various exact solutions can be constructed. Specifically, the Riccati equation admits the following forms:
\begin{align}\label{5}
\zeta =
\begin{cases}
-\dfrac{1}{2} \delta \left[1 + \coth\left( \dfrac{1}{2} \delta q \right)\right], & \delta > 0, \\
-\dfrac{1}{2} \delta \left[1 + \tanh\left( \dfrac{1}{2} \delta q \right)\right], & \delta < 0, \\
-\dfrac{1}{q}, & \delta = 0.
\end{cases}
\end{align}

By substituting Eq.~\eqref{5} into the Eq. \eqref{4} and employing balancing techniques, explicit expressions for the coefficient functions $A$, $B_i$, $C_i$, and the function $q(x,y,t)$ were derived.

Upon application to the BLP system, from Eq. \eqref{4} this method yielded simplified expressions for $u$ and $v$:
\begin{align}
u &= a_1(x,y,t) + b_1(x,y,t)\zeta(q(x,y,t)) + c_1(x,y,t) \sqrt{\delta \zeta(q) + \zeta^2(q)}, \\
v &= a_2(x,y,t) + b_2(x,y,t)\zeta(q(x,y,t)) + c_2(x,y,t) \sqrt{\delta \zeta(q) + \zeta^2(q)},
\end{align}
where $a_1, b_1, c_1, a_2, b_2, c_2$, and $q$ are all regarded as functions of $(x, y, t)$. These were identified by substituting into the governing system Eq. \eqref{1} and equating coefficients of like powers of $\zeta$ to zero, yielding the following:
\begin{align}
a_1 &= -\dfrac{1}{4} \dfrac{-2q_t + 2q_{xx} + \delta q_x^2}{q_x}, \quad b_1 = -\dfrac{1}{2}q_x, \quad c_1 = \dfrac{1}{2}q_x, \nonumber \\
a_2 &= -\dfrac{1}{2}q_y, \quad b_2 = \dfrac{1}{2}q_y, \nonumber \\
c_2 &= -\dfrac{1}{4} \left[ q_x^3 q_{xy} \delta - 2q_x q_{xy} q_{xxx} + 2q_x q_{xxy} q_t + 2q_x q_{xy} q_{xt} + 2q_{xxxy} q_x^2 - 2q_{xy} q_x^2 
\right. \nonumber \\
& \quad \left. + 2q_{xy} q_t q_x + 4q_{xy} q_{xx} - 4q_{xx} q_{xy} q_t - 4q_{xx} q_{xxy} q_x \right] / q_x^2.
\end{align}
Separation of variables is now applied to the function $q(x,y,t)$, which is represented as
\begin{equation}
q(x,y,t) = \psi(x,t) + \phi(y),
\end{equation}
where $\psi(x,t)$ and $\phi(y)$ are considered as arbitrary functions of the variables $x, t$ and $y$, respectively.
From these expressions, explicit solutions of the BLP system were derived:
\begin{align}
u_1 &= \dfrac{1}{2} \dfrac{-\psi_{xx} + \psi_t}{\psi_x} +
\dfrac{1}{4} \psi_x \delta \left\{ \tanh\left( \dfrac{1}{2} \delta (\psi + \phi) \right)
\sqrt{\tanh^2\left( \dfrac{1}{2} \delta (\psi + \phi) \right)-1} \right\}, \\
v_1 &= \dfrac{1}{4} \delta \phi_y \left\{ 1 + \tanh\left( \dfrac{1}{2} \delta (\psi + \phi) \right)
\sqrt{ \tanh^2\left( \dfrac{1}{2} \delta (\psi + \phi) \right)-1} \right\}, \\
u_2 &= \dfrac{1}{2} \dfrac{-\psi_{xx} + \psi_t}{\psi_x} +
\dfrac{1}{4} \psi_x \delta \left\{ \coth\left( \dfrac{1}{2} \delta (\psi + \phi) \right)
\sqrt{ \coth^2\left( \dfrac{1}{2} \delta (\psi + \phi) \right)-1} \right\},\label{12} \\
v_2 &= \dfrac{1}{4} \delta \phi_y \left\{ 1 + \coth\left( \dfrac{1}{2} \delta (\psi + \phi) \right)
\sqrt{ \coth^2\left( \dfrac{1}{2} \delta (\psi + \phi) \right)-1} \right\}.\label{13}
\end{align}

For the arbitrariness of the functions $\psi(x,t)$ and $\phi(y)$, the solution space is significantly enriched. By differentiating Eq.~\eqref{12} with respect to $y$, similarly differentiating Eq.~\eqref{13} with respect to $x$, the unique gradient function $U$ is obtained as follows:
\begin{equation}\label{14}
U = u_{2y} = v_{2x} = -\dfrac{1}{8} \psi_x \phi_y \delta^2 \ \text{cosech} \left( \dfrac{1}{2} \delta (x + \phi) \right)
\left[ \text{cosech} \left( \dfrac{1}{2} \delta (x + \phi) \right) + \coth \left( \dfrac{1}{2} \delta (x + \phi) \right) \right].
\end{equation}

This set of explicit solutions reveals complex nonlinear dynamics within the BLP system, including localized waveforms and intricate spatial features, laying the groundwork for further fractal analysis.
\section{Fractal structures:}\label{S3}
The formation of fractal structures in nonlinear systems is strongly driven by the scale-sensitive and oscillatory properties of the functions involved. In this work, three distinct function types are considered. Type-I uses nested trigonometric-logarithmic forms like $\sin(\log(\cdot))$ to induce multi-scale oscillations. Type-II employs Jacobi elliptic functions (e.g., \texttt{JacobiSN}, \texttt{JacobiDN}) to model quasi-periodic patterns typical in bounded wave systems. Type-III blends rational envelopes with $\sin^2(\log(\cdot))$ and $\cos^2(\log(\cdot))$ components to produce localized yet richly oscillatory structures. Each formulation leads to self-similar figures that reveal finer details upon magnification, underscoring the role of functional choice in fractal emergence.

The type-I function, defined below,
\begin{equation}\label{15}
\left.
\begin{array}{l}
\psi = 1 + \exp\left( -(x+t) \left( (x+t) + \cos\left(\log((x+t)^2)\right) 
+ \sin\left(\log((x+t)^2 )\right) \right) \right), \\[1ex]
\phi = 1 + \exp\left( -y \left( y + \cos\left(\log(y^2 )\right) 
+ \sin\left(\log(y^2 )\right) \right) \right),
\end{array}
\right\}
\end{equation}
is incorporated into Eq.~\eqref{14} to generate the target structures. \par
\begin{figure}[h]
    \centering

\begin{subfigure}[b]{0.48\textwidth}
    \includegraphics[width=\linewidth]{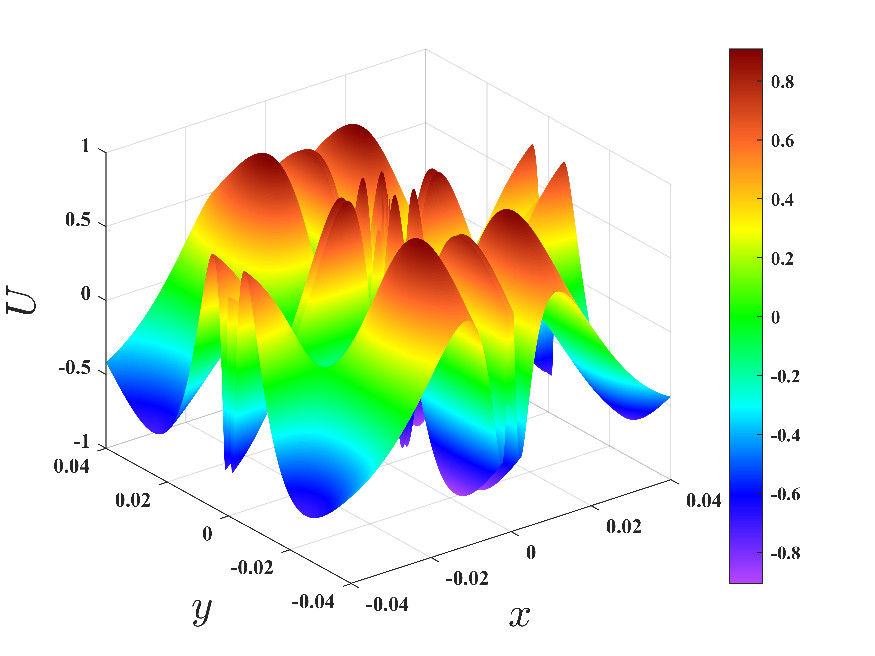}
    \caption{(a)}
    \label{f1a}
\end{subfigure}
\hfill
\begin{subfigure}[b]{0.45\textwidth}
    \includegraphics[width=\linewidth]{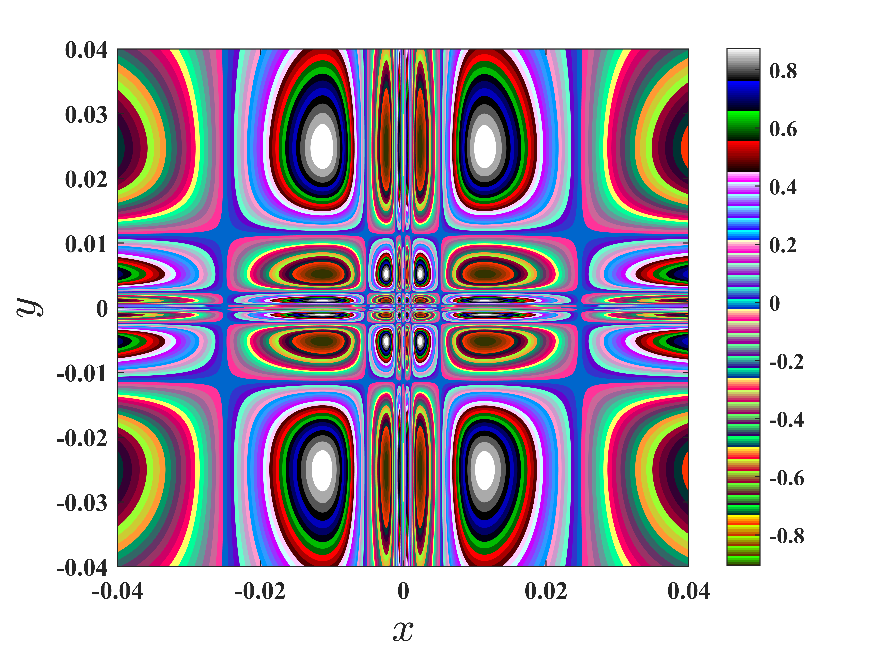}
    \caption{(b)}
    \label{f1b}
\end{subfigure}

\vspace{0.2em}

\begin{subfigure}[b]{0.48\textwidth}
    \includegraphics[width=\linewidth]{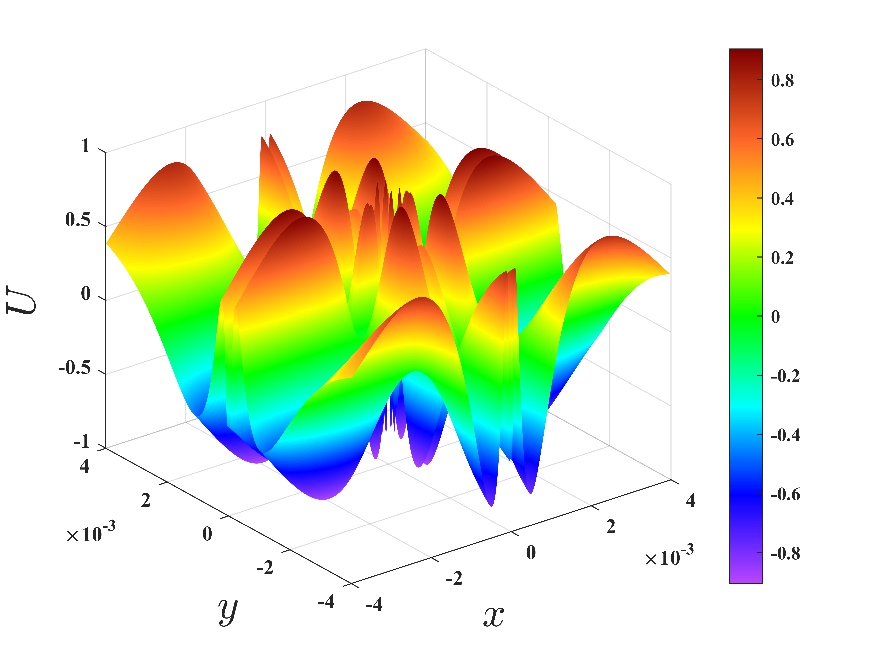}
    \caption{(c)}
    \label{f1c}
\end{subfigure}
\hfill
\begin{subfigure}[b]{0.45\textwidth}
    \includegraphics[width=\linewidth]{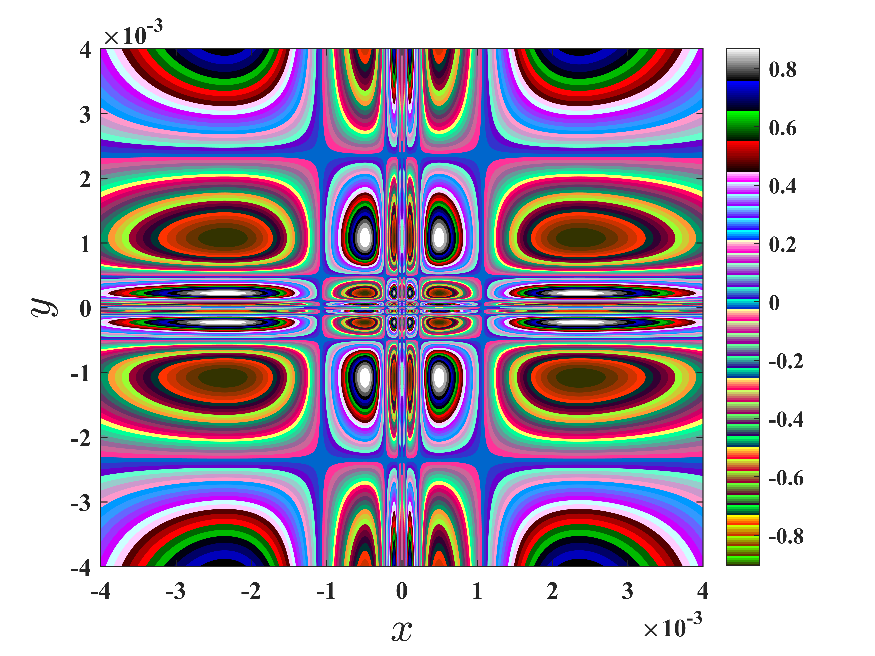}
    \caption{(d)}
    \label{f1d}
\end{subfigure}

\vspace{0.2em}

\begin{subfigure}[b]{0.48\textwidth}
    \includegraphics[width=\linewidth]{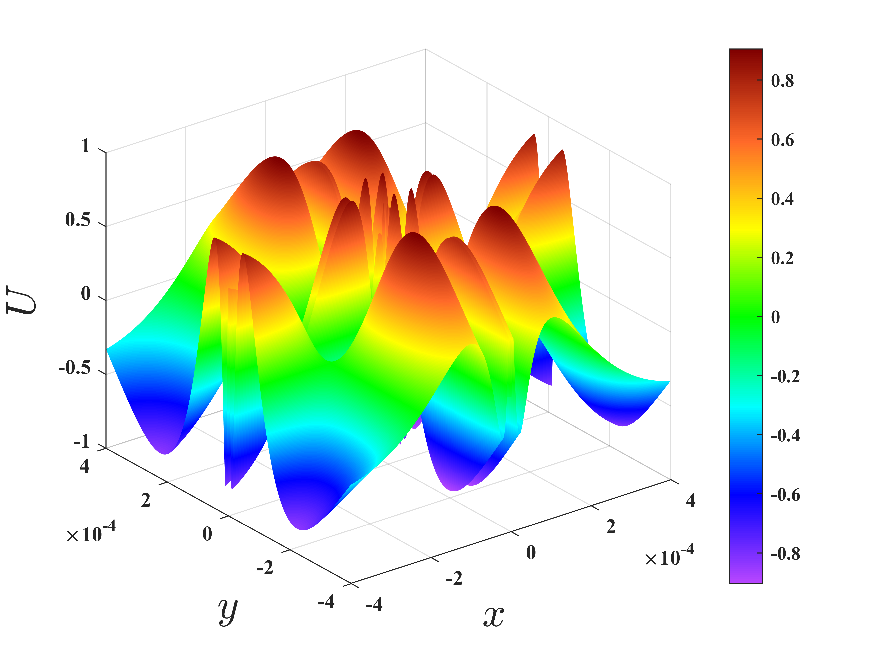}
    \caption{(e)}
    \label{f1e}
\end{subfigure}
\hfill
\begin{subfigure}[b]{0.45\textwidth}
    \includegraphics[width=\linewidth]{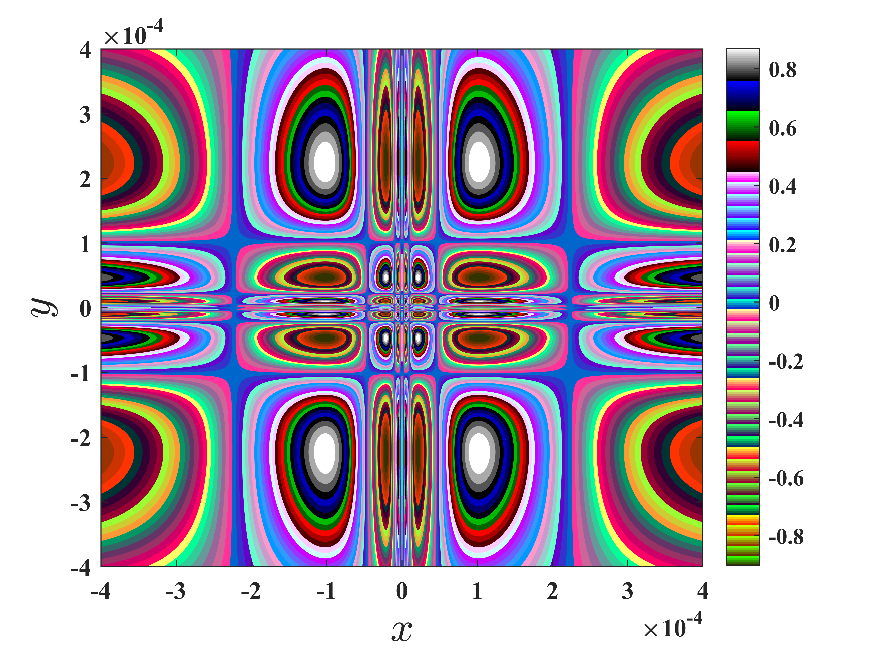}
    \caption{(f)}
    \label{f1f}
\end{subfigure}

    \caption{A progressive zoom into the three-dimensional fractal structures and their associated contour profiles, governed by Eq.~\eqref{14} and constructed using Eq.~\eqref{15} with the parameter set $\delta=1$ and $t = 0$, is presented. Supplementary videos corresponding to \figurename~\ref{f1} are available: {\href{https://drive.google.com/file/d/1sgpOmEkYumySswY6zgamSjrmvPo9hsrr/view?usp=sharing} {{\color{blue} three-dimensional fractal structures}}} and {\href{https://drive.google.com/file/d/1ewR8tk79JZnWLi2tTrJN2ZwiCMiEPVPL/view?usp=sharing}{\color{blue} corresponding contour profiles}}.
}
    \label{f1}
\end{figure}

\figurename~\ref{f1} illustrates a sequential zoom-in of the three-dimensional fractal structures and their corresponding contour plots, derived from Eq.~\eqref{14} using functions defined in Eq.~\eqref{15}, under the fixed parameters $t=1$ and $\mu=1$. The \figurename s~\ref{f1a}–\ref{f1f} demonstrate how the structure evolves across scales, with each level of magnification revealing finer, self-similar patterns—a hallmark of fractal geometry. As the domain narrows, both the surface and contour plots exhibit increasingly intricate waveforms while preserving the overall structure. This consistent recurrence of geometric features across zoom levels confirms the fractal nature embedded in the functional formulation.\par
The function referred to as type-II is introduced below
\begin{equation}\label{16}
\left.
\begin{array}{l}
	\psi = 1 + e^{ 
		1 - (x + t) \left( (x + t) + \text{JacobiDN} \left( \sin^2 \left( \log{x^2} \right), 0.7 \right)  
		+ \text{JacobiSN} \left( \cos^2 \left( \log{x^2} \right), 0.7 \right) \right) }, \\[10pt]
	\phi = 1 + e^{ 
		1 - (y + t) \left( (y + t) + \text{JacobiDN} \left( \sin^2 \left( \log{y^2} \right), 0.7 \right)  
		+ \text{JacobiSN} \left( \cos^2 \left( \log{y^2} \right), 0.7 \right) \right) },
\end{array} \right\}
\end{equation}
and applied within Eq.~\eqref{14} to obtain the desired fractal patterns.
\par

\begin{figure}[h]
    \centering

\begin{subfigure}[b]{0.48\textwidth}
    \includegraphics[width=\linewidth]{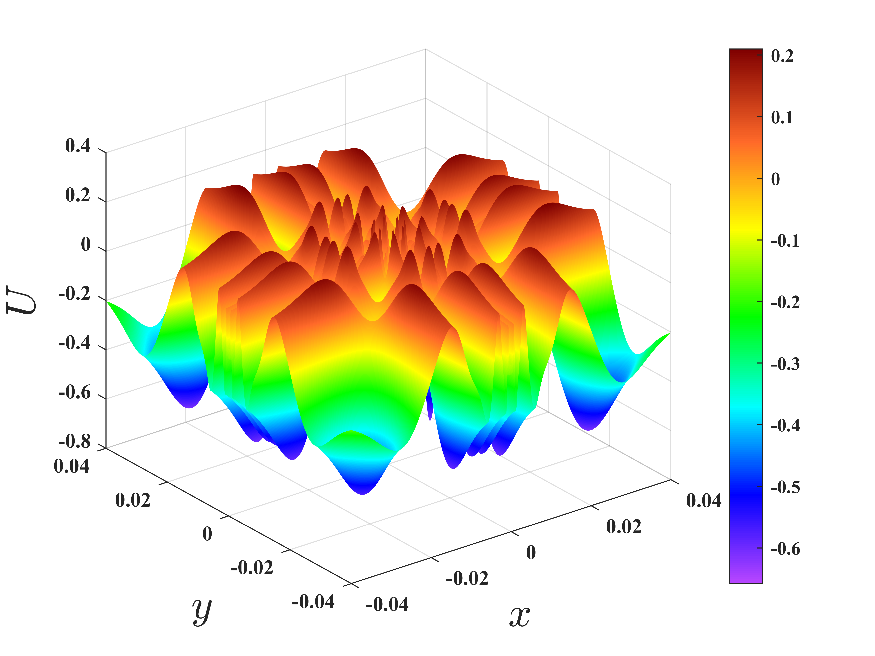}
    \caption{(a)}
    \label{f2a}
\end{subfigure}
\hfill
\begin{subfigure}[b]{0.45\textwidth}
    \includegraphics[width=\linewidth]{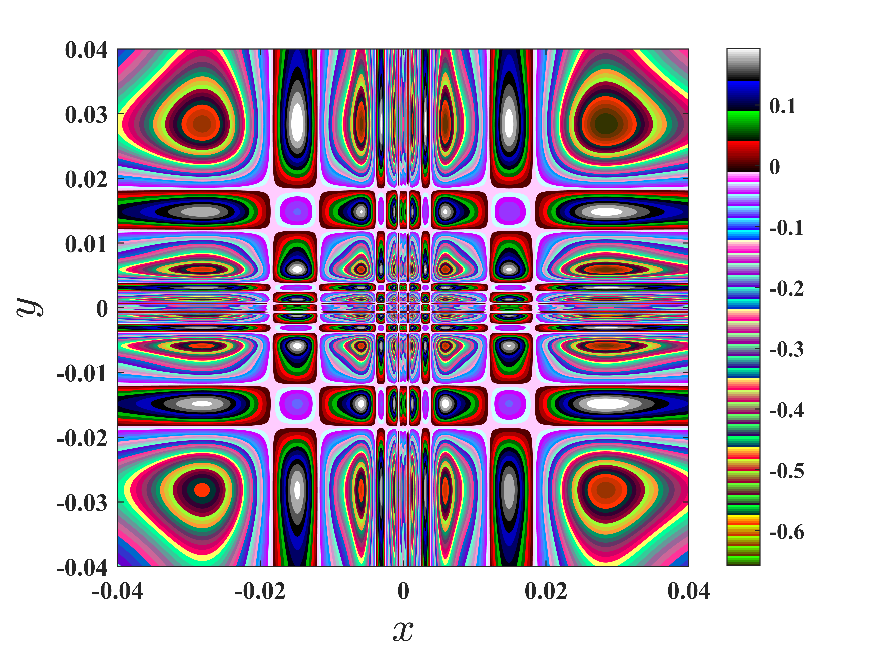}
    \caption{(b)}
    \label{f2b}
\end{subfigure}

\vspace{0.2em}

\begin{subfigure}[b]{0.48\textwidth}
    \includegraphics[width=\linewidth]{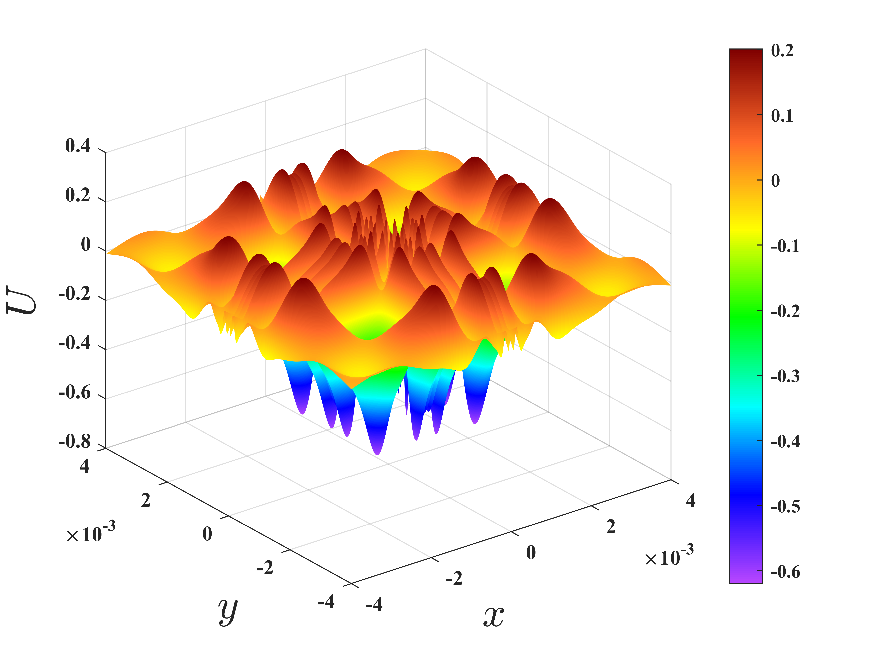}
    \caption{(c)}
    \label{f2c}
\end{subfigure}
\hfill
\begin{subfigure}[b]{0.45\textwidth}
    \includegraphics[width=\linewidth]{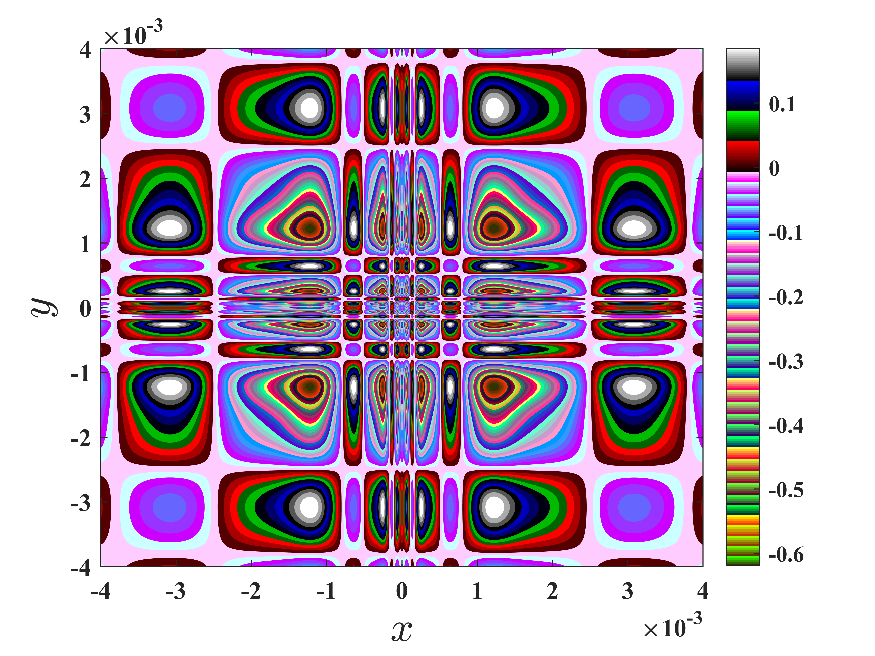}
    \caption{(d)}
    \label{f2d}
\end{subfigure}

\vspace{0.2em}

\begin{subfigure}[b]{0.48\textwidth}
    \includegraphics[width=\linewidth]{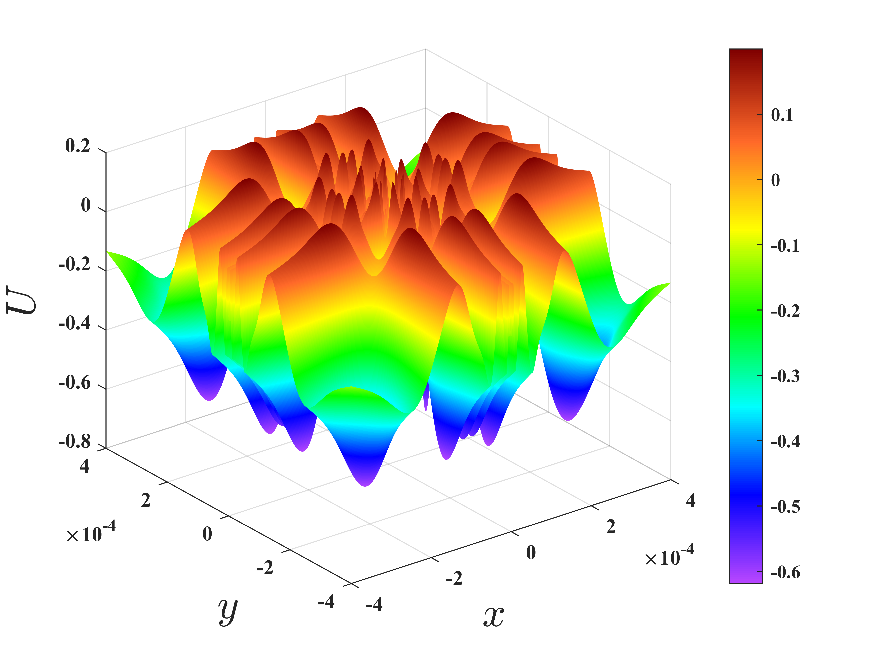}
    \caption{(e)}
    \label{f2e}
\end{subfigure}
\hfill
\begin{subfigure}[b]{0.45\textwidth}
    \includegraphics[width=\linewidth]{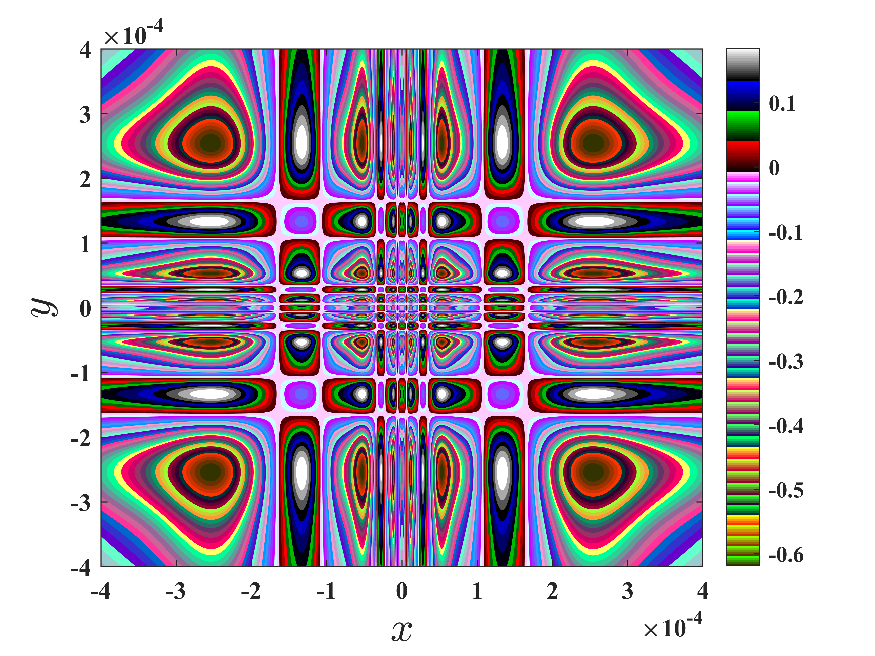}
    \caption{(f)}
    \label{f2f}
\end{subfigure}
    \caption{A progressive zoom into the three-dimensional fractal structures and their associated contour profiles, governed by Eq.~\eqref{14} and constructed using Eq.~\eqref{16} with the parameter set $\delta=1$ and $t = 0$, is presented.  Supplementary videos corresponding to \figurename~\ref{f2} are available: \href{https://drive.google.com/file/d/1kdt3E4m2U-akxJGpiv3jBGj5sBWM6xbU/view?usp=sharing}{{\color{blue}three-dimensional fractal structures}} and \href{https://drive.google.com/file/d/1-owF1UQGADMxHv04iE42rOp7zD7iO-nu/view?usp=sharing}{{\color{blue} corresponding contour profiles}}.}
    \label{f2}
\end{figure}

\figurename~\ref{f2} presents successive magnifications of three-dimensional fractal profiles and their associated contour plots, generated from Eq.~\eqref{14} using the expressions in Eq.~\eqref{16}, with fixed parameters $t = 1$ and $\delta= 1$. \figurename s~\ref{f2a}–\ref{f2f} reveal consistent self-similar structures across multiple scales, confirming the scale-invariant nature of the solution. As the zoom level increases, finer wave patterns and sharper peaks emerge in both the surface and contour plots, while preserving the overall geometry. Each zoom stage exhibits recurring motifs, emphasizing the fractal characteristics embedded in the functional construction. These results further validate the recursive and structured complexity of the underlying solution.
\par

Again, the function categorized as type-III is defined below 
\begin{equation}\label{17} 
\left.
\begin{array}{l}
	\psi = 1 + \left( \frac{x + t}{1 + (x + t)^4} \right) \sin^2 \left( \log \left( (x + t)^2 \right) \right) \\[10pt]
	\phi = 1 + \left( \frac{y }{1 + y^4} \right) \cos^2 \left( \log \left( y ^2 \right) \right)
    \end{array}
    \right\}
\end{equation}
and is employed in Eq.~\eqref{14} to construct the corresponding fractal structures. 

\begin{figure}[h]
    \centering

\begin{subfigure}[b]{0.48\textwidth}
    \includegraphics[width=\linewidth]{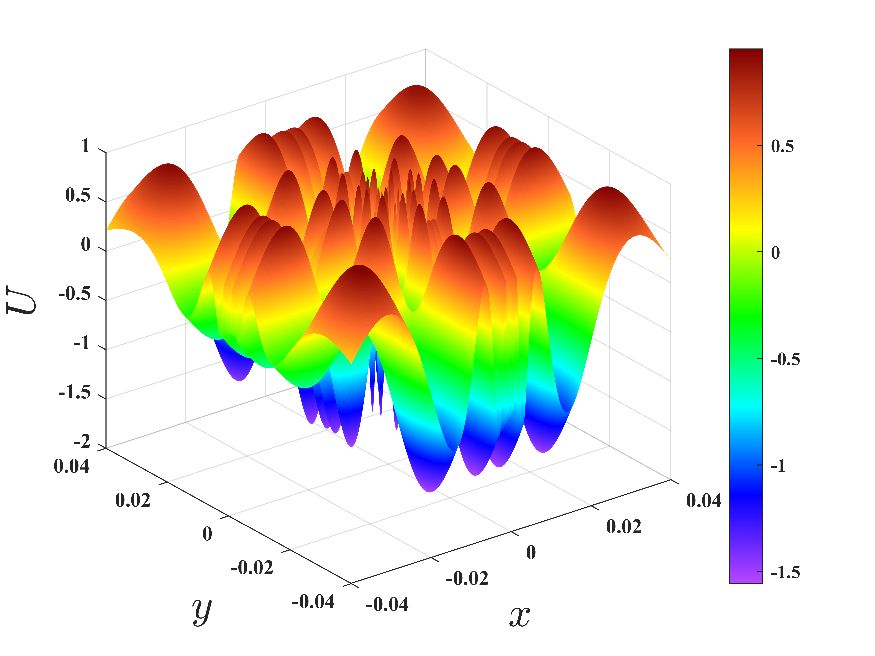}
    \caption{(a)}
    \label{f3a}
\end{subfigure}
\hfill
\begin{subfigure}[b]{0.45\textwidth}
    \includegraphics[width=\linewidth]{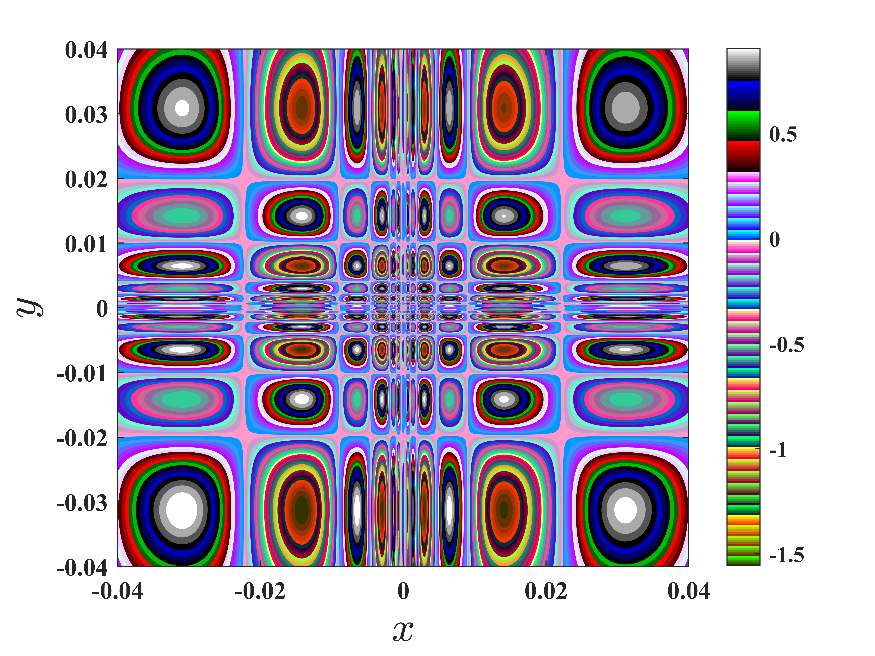}
    \caption{(b)}
    \label{f3b}
\end{subfigure}

\vspace{0.2em}

\begin{subfigure}[b]{0.48\textwidth}
    \includegraphics[width=\linewidth]{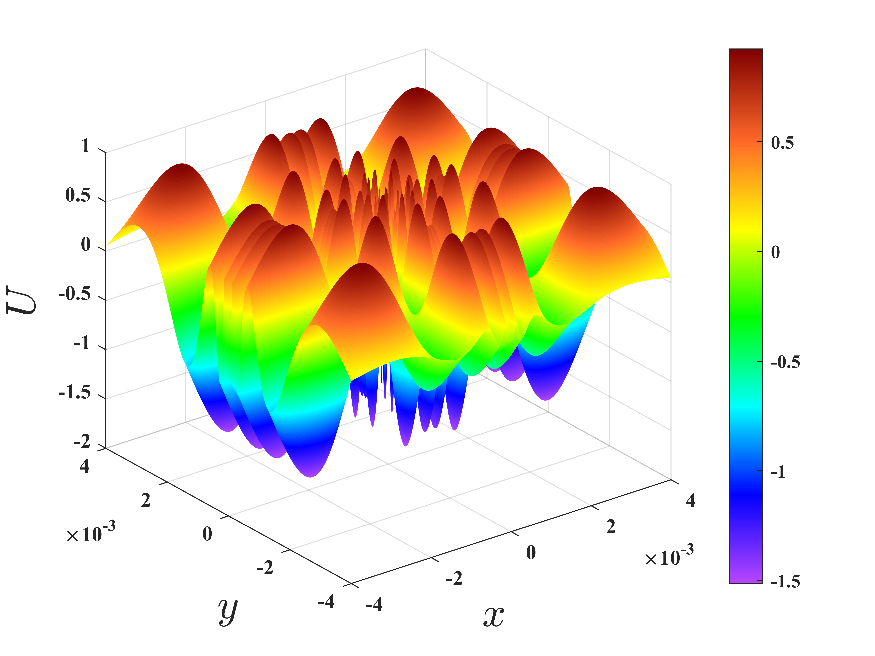}
    \caption{(c)}
    \label{f3c}
\end{subfigure}
\hfill
\begin{subfigure}[b]{0.45\textwidth}
    \includegraphics[width=\linewidth]{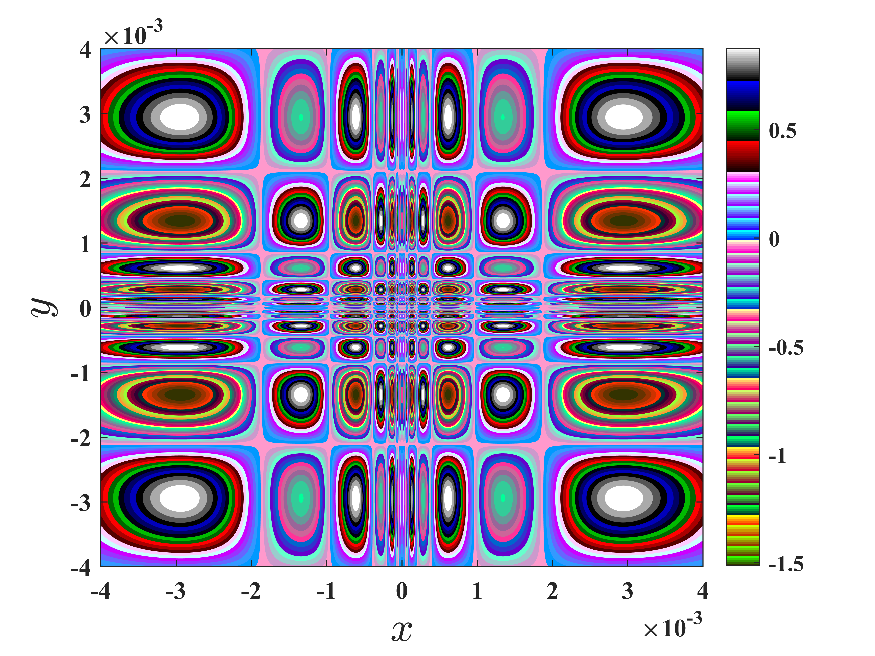}
    \caption{(d)}
    \label{f3d}
\end{subfigure}

\vspace{0.2em}

\begin{subfigure}[b]{0.48\textwidth}
    \includegraphics[width=\linewidth]{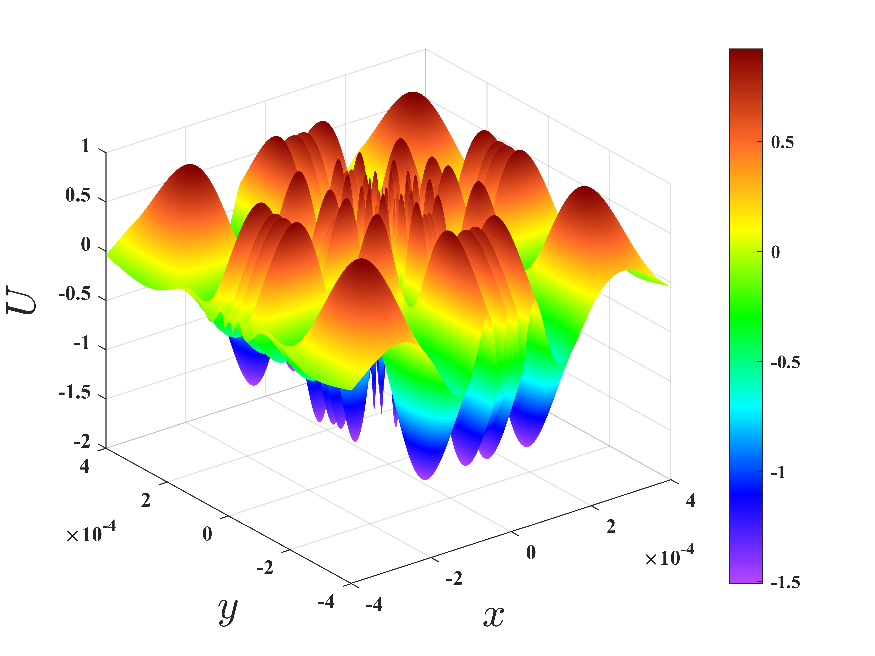}
    \caption{(e)}
    \label{f3e}
\end{subfigure}
\hfill
\begin{subfigure}[b]{0.45\textwidth}
    \includegraphics[width=\linewidth]{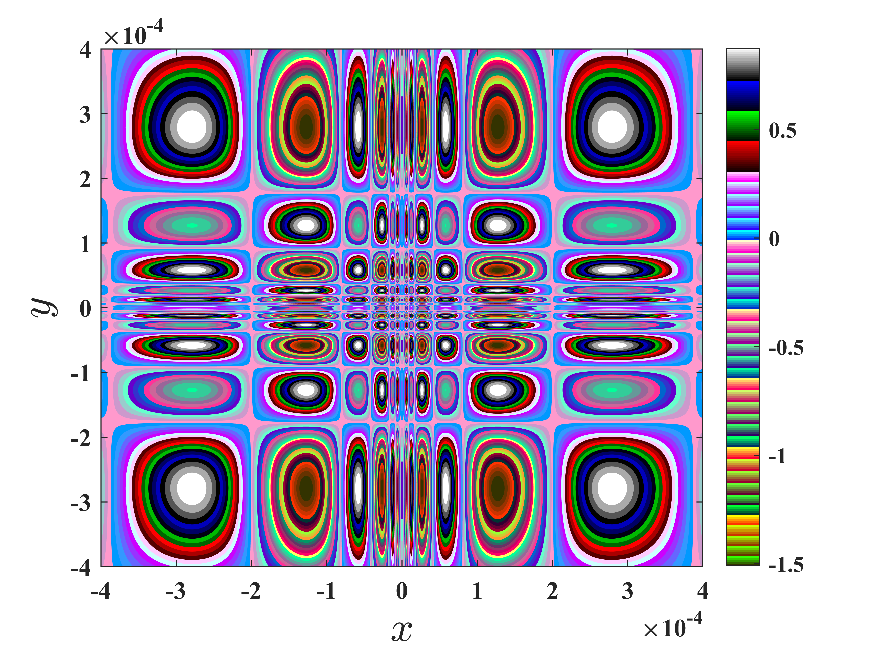}
    \caption{(f)}
    \label{f3f}
\end{subfigure}

    \caption{A progressive zoom into the three-dimensional fractal structures and their associated contour profiles, governed by Eq.~\eqref{14} and constructed using Eq.~\eqref{17} with the parameter set $\delta=1$ and $t = 0$, is presented. Supplementary videos corresponding to \figurename~\ref{f3} are available: \href{https://drive.google.com/file/d/11014CifNeWaV4D1p1drzMcyYL-AWaBsY/view?usp=sharing}{{\color{blue}three-dimensional fractal structures}} and \href{https://drive.google.com/file/d/1YrVcFzbXlnIMuzBS8-Q5qMKJg4q35cJi/view?usp=sharing}{{\color{blue} corresponding contour profiles}}.}
    \label{f3}
\end{figure}

\figurename~\ref{f3} illustrates staged magnifications of three-dimensional fractal waveforms and their corresponding contour plots, derived from Eq.~\eqref{14} using the formulations in Eq.~\eqref{17}, with the same parameter set as in \figurename s~\ref{f1} and \ref{f2}. \figurename s~\ref{f3a}–\ref{f3f} reveal increasingly intricate structures as the view narrows, highlighting the recursive and self-similar nature of the solution. The consistent appearance of peaks, troughs, and symmetric contour patterns across scales underscores the scale-invariant properties inherent in fractal systems. Notably, each zoom level preserves the overall geometric framework, with sharper gradients and finer oscillations emerging at higher resolutions. These results demonstrate the fractal behavior encoded in the functional construction of the solution.
\clearpage
\section{Fractal dimension:}\label{S4}
By selecting different functional forms for $\psi(x,y)$ and $\phi(t)$ in Eq.~\eqref{14}, various fractal surface structures have been generated, as shown in \figurename s~\ref{f1}–\ref{f3}. Each subfigure presents the three-dimensional fractal surfaces (left) and their corresponding contour plots (right), highlighting self-similar patterns across multiple scales. Fractals, characterized by non-integer dimensions and recursive geometry, offer a robust framework for analyzing structural complexity. To quantify this, the box-counting dimension is employed, defined as
\[
\dim_B(A) = \lim_{\varepsilon \to 0} \frac{\log N_\varepsilon(A)}{\log(1/\varepsilon)},
\]
where $N_\varepsilon(A)$ denotes the number of boxes of size $\varepsilon$ needed to cover the set $A$. In this study, it is estimated by plotting $\log N(\varepsilon)$ versus $\log(1/\varepsilon)$ and computing the slope. The resulting non-integer dimensions confirm the self-similar and intricate nature of the BLP system's fractal solutions.

\begin{figure}[h]
    \centering

    \begin{subfigure}[b]{0.32\textwidth}
        \includegraphics[width=\linewidth]{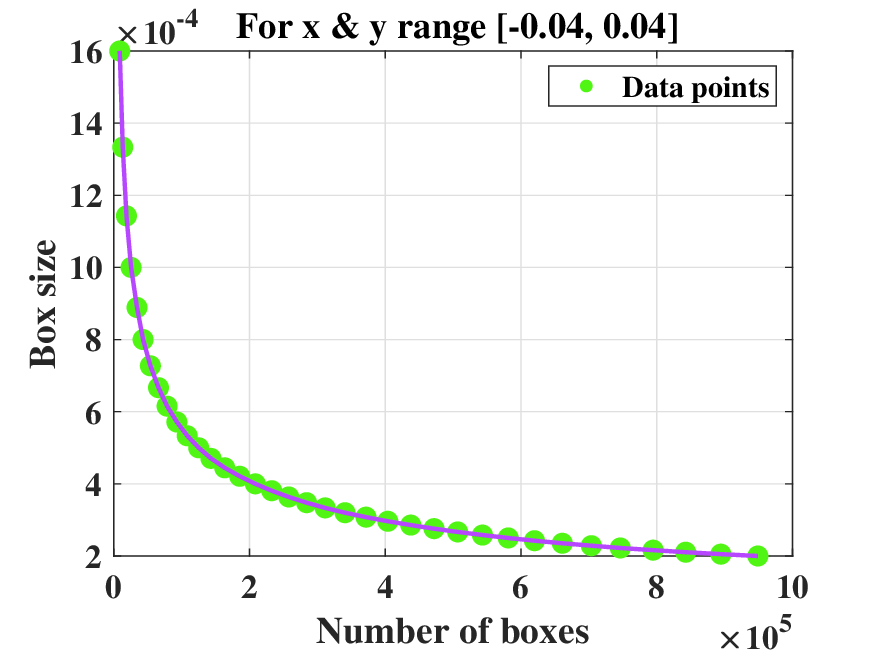}
        \caption{}
    \end{subfigure}
    \begin{subfigure}[b]{0.32\textwidth}
        \includegraphics[width=\linewidth]{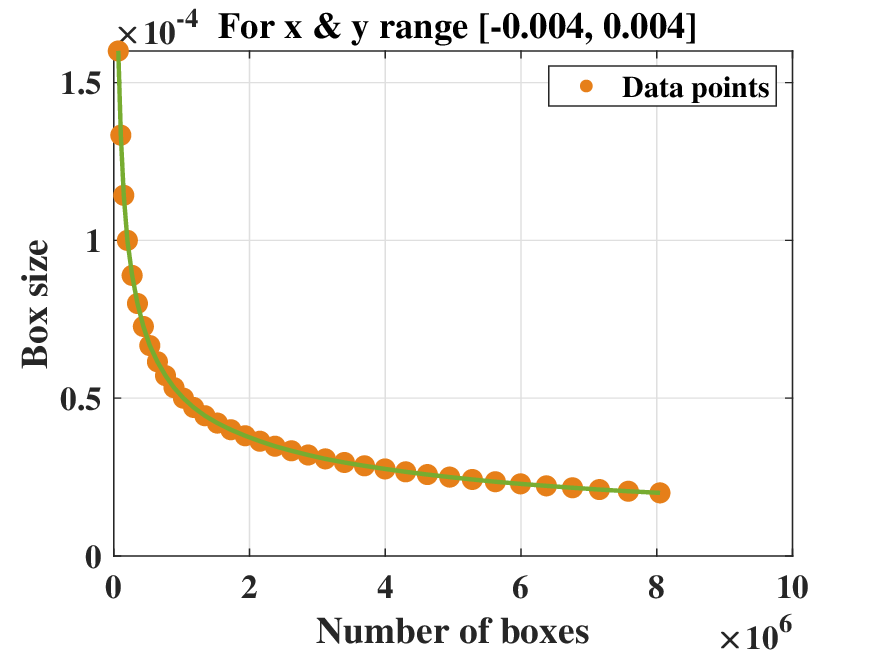}
        \caption{}
    \end{subfigure}
    \begin{subfigure}[b]{0.32\textwidth}
        \includegraphics[width=\linewidth]{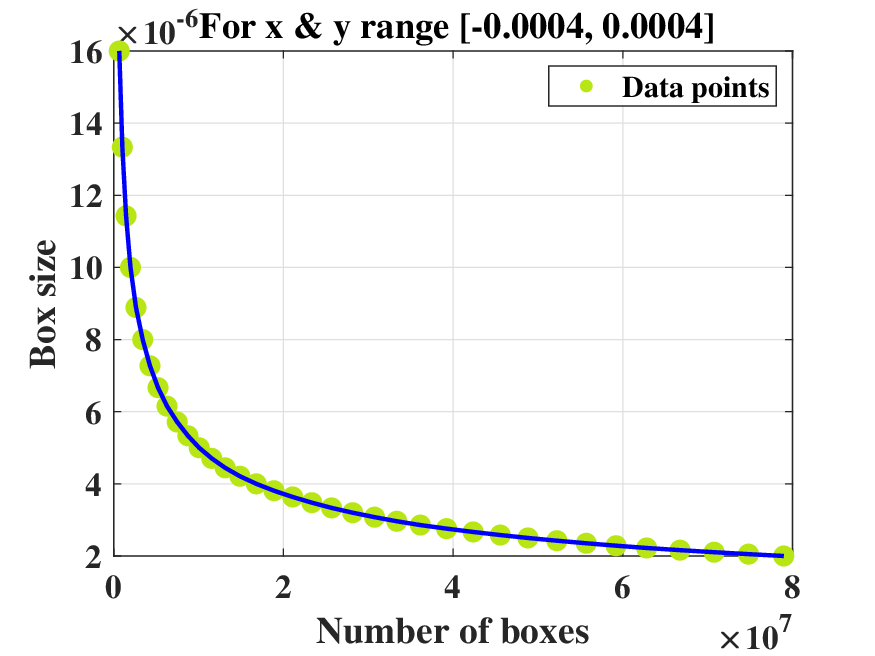}
        \caption{}
    \end{subfigure}

    \caption{Two-dimensional plots of box size \( (\varepsilon) \) versus number of boxes \(( N(\varepsilon)) \) covering the fractal surface from \figurename~\ref{f1} at three zoom levels in \( x \) and \( y \). Data points represent computed box-counting results. }
    \label{f4}
\end{figure}
\begin{figure}[h]
    \centering

    \begin{subfigure}[b]{0.32\textwidth}
        \includegraphics[width=\linewidth]{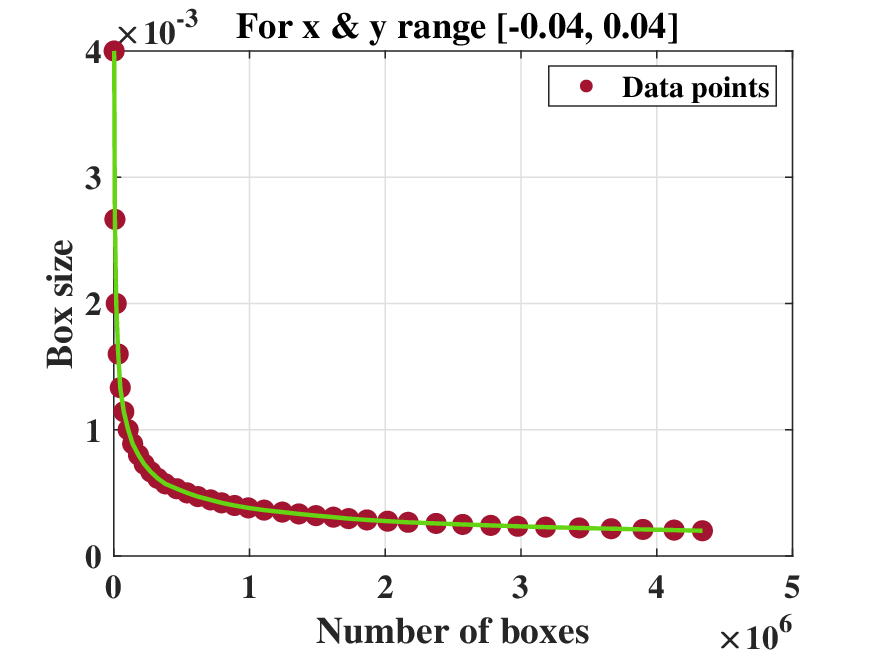}
        \caption{}
    \end{subfigure}
    \begin{subfigure}[b]{0.32\textwidth}
        \includegraphics[width=\linewidth]{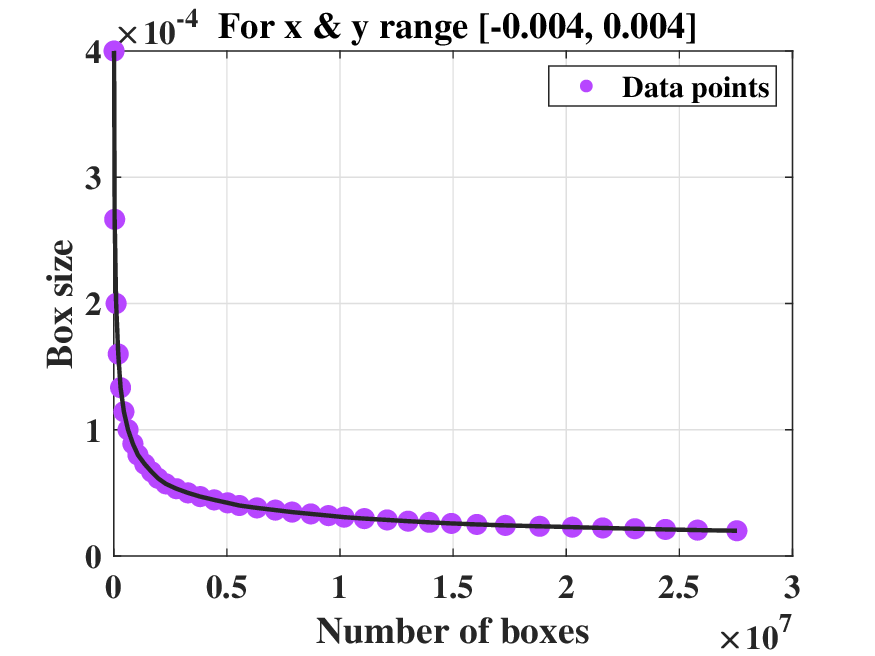}
        \caption{}
    \end{subfigure}
    \begin{subfigure}[b]{0.32\textwidth}
        \includegraphics[width=\linewidth]{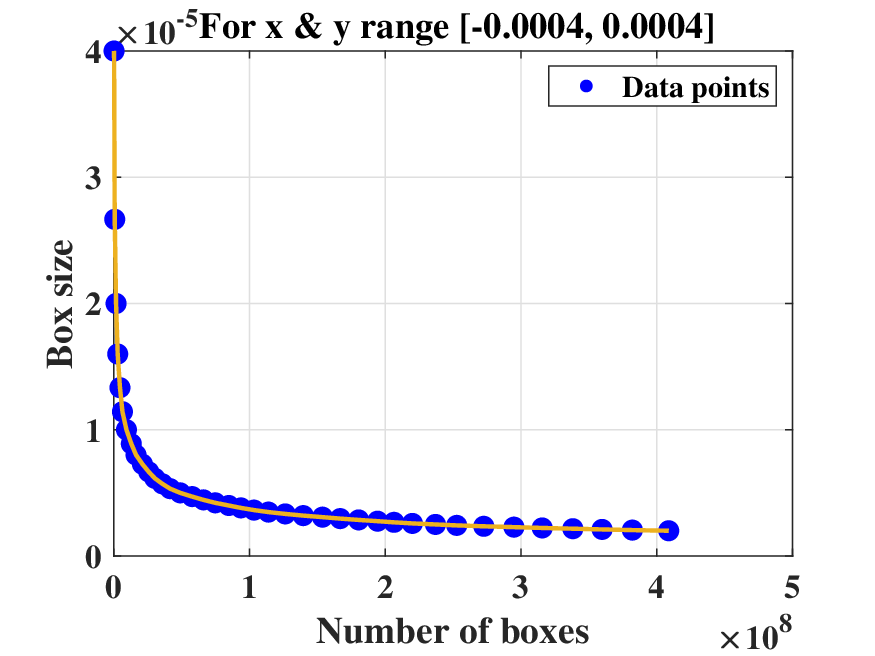}
        \caption{}
    \end{subfigure}

    \caption{Two-dimensional plots of box size \((\varepsilon)\) versus number of boxes \( (N(\varepsilon)) \) covering the fractal surface from \figurename~\ref{f2} at three zoom levels in \( x \) and \( y \). Data points represent computed box-counting results. }
    \label{f5}
\end{figure}

\begin{figure}[h]
    \centering

    \begin{subfigure}[b]{0.32\textwidth}
        \includegraphics[width=\linewidth]{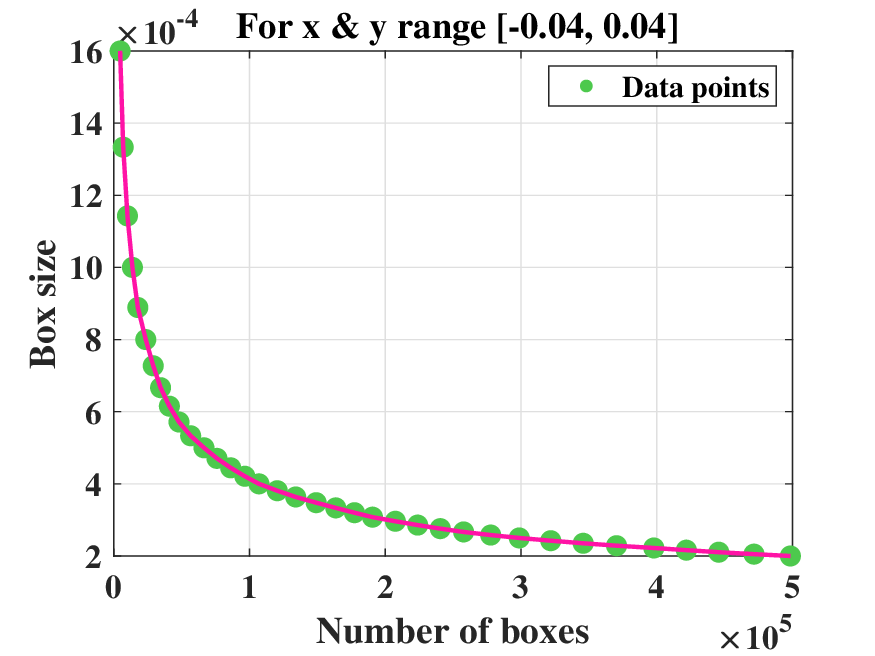}
        \caption{}
    \end{subfigure}
    \begin{subfigure}[b]{0.32\textwidth}
        \includegraphics[width=\linewidth]{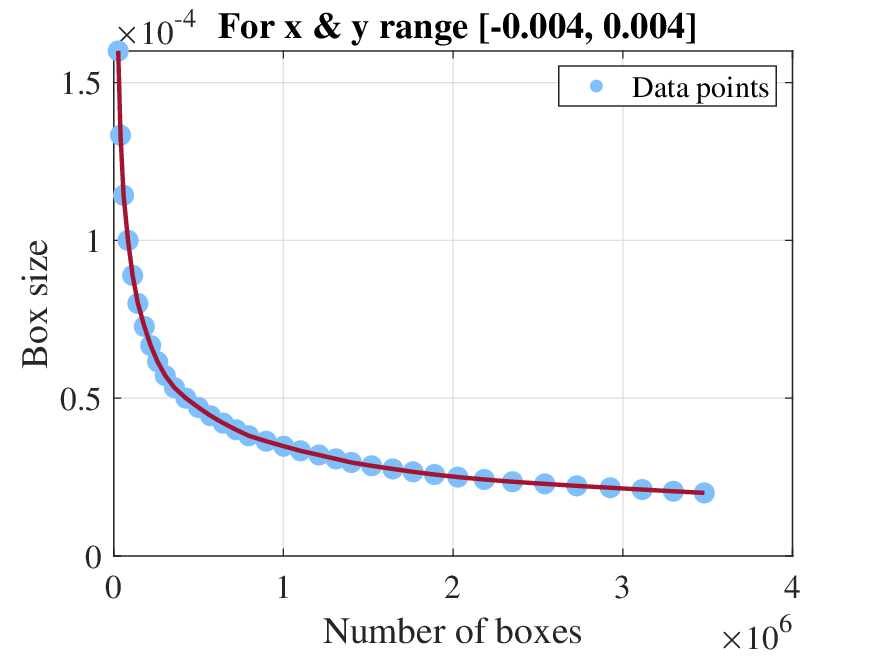}
        \caption{}
    \end{subfigure}
    \begin{subfigure}[b]{0.32\textwidth}
        \includegraphics[width=\linewidth]{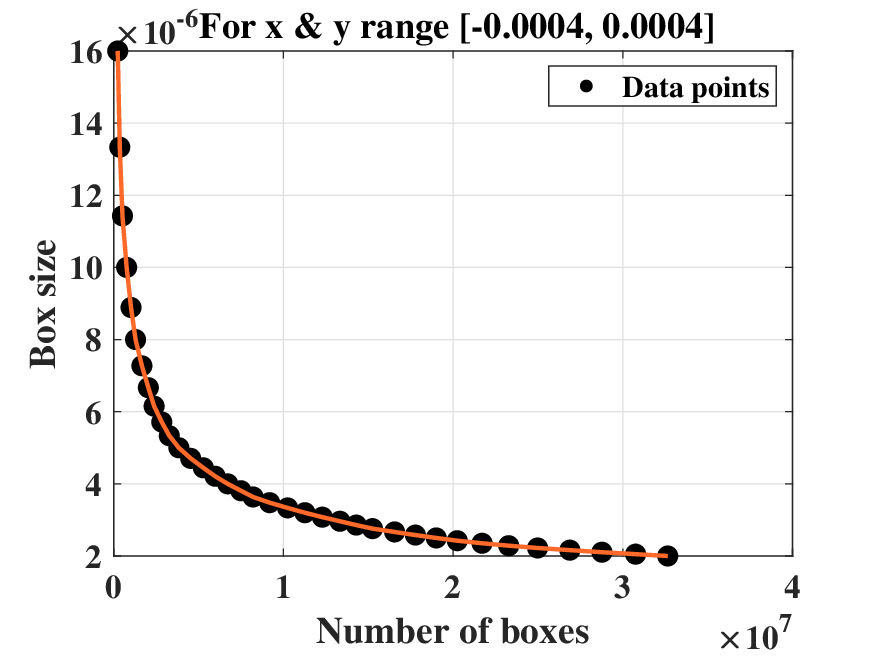}
        \caption{}
    \end{subfigure}

    \caption{Two-dimensional plots of box size \(( \varepsilon) \) versus number of boxes \( (N(\varepsilon) )\) covering the fractal surface from \figurename~\ref{f3} at three zoom levels in \( x \) and \( y \). Data points represent computed box-counting results. }
    \label{f6}
\end{figure}
\figurename s~\ref{f4}–\ref{f6} display the results of a voxel-based box-counting analysis applied to fractal surface profiles derived from the BLP system. Each figure showcases a sequence of zoomed-in surface views over progressively smaller domains in the $x$--$y$ plane, revealing increasing geometric intricacy at finer spatial scales. As resolution improves, the voxel count needed to cover the surface increases in a nonlinear manner, indicating the presence of self-similar and complex structures. Circular markers highlight the data points extracted from box-counting plots, representing the relationship between box size and the number of boxes required to cover the surface. These markers indicate sampling positions used in the dimension estimation. Collectively, the figures offer compelling visual evidence of fractal scaling behavior within the BLP system, reinforcing the interpretation of its solutions as structurally self-similar across multiple magnification levels.

\begin{table}[h]
\centering
\caption{Box-counting dimension estimates corresponding to \figurename s~\ref{f1}--\ref{f3} at various spatial domains in the $x$ and $y$ directions.}
\scriptsize 
\begin{tabular}{ccc}
\begin{tabular}{ll}
\toprule
\textbf{$x$ \& $y$ Range} & \textbf{Dim} \\
\midrule
$[-0.04,0.04]$ & 1.6159 \\
$[-0.004,0.004]$ & 1.4696 \\
$[-0.0004,0.0004]$ & 1.3858 \\
\bottomrule
\end{tabular}
&
\begin{tabular}{ll}
\toprule
\textbf{$x$ \& $y$ Range} & \textbf{Dim} \\
\midrule
$[-0.04,0.04]$ & 1.7943 \\
$[-0.004,0.004]$ & 1.5833 \\
$[-0.0004,0.0004]$ & 1.5110 \\
\bottomrule
\end{tabular}
&
\begin{tabular}{ll}
\toprule
\textbf{$x$ \& $y$ Range} & \textbf{Dim} \\
\midrule
$[-0.04,0.04]$ & 1.5403\\
$[-0.004,0.004]$ & 1.3921 \\
$[-0.0004,0.0004]$ & 1.3185 \\
\bottomrule
\end{tabular}
\vspace{0.6em}
\\
(a) For \figurename~\ref{f1} & (b) For \figurename~\ref{f2} & (c) For \figurename~\ref{f3}
\end{tabular}
\label{tab1}
\end{table}

Table~\ref{tab1} presents the box-counting dimension values calculated from the visual analyses shown in \figurename s~\ref{f4}–\ref{f6}, which are linked to the fractal surfaces depicted in \figurename s~\ref{f1}–\ref{f3} across three levels of zoom. As the spatial domain $(x, y)$ is progressively reduced, the estimated fractal dimension tends to decrease, suggesting a shift from highly intricate to relatively simpler structures. These findings support the existence of non-integer dimensionality, a hallmark of fractal geometry. This scale-dependent variation in complexity provides further insight into the local structural behavior of the system.

\section{Conclusion:}\label{S5}

This research presents a comprehensive investigation of fractal structures arising in the solutions of the Boiti–Leon–Pempinelli system. By incorporating carefully chosen oscillatory functions into the analytical framework, the study explores how nonlinear interactions within the BLP system give rise to complex, scale-invariant patterns. The analysis combines qualitative visualization with quantitative dimension estimation, revealing key aspects of the system's geometric intricacy. The major findings are outlined below:

\begin{enumerate}
    \item[a.] Three types of analytical functions—nested trigonometric-logarithmic expressions, Jacobi elliptic forms, and rationally modulated oscillations—were employed to construct solutions of the BLP system. These functions led to the emergence of detailed, self-similar surface profiles.

    \item[b.] Successive zoom-in visualizations of the three-dimensional structures and their corresponding contour plots demonstrated the recursive, fractal-like nature of the solutions. Repeating patterns appeared consistently across different spatial scales.

    \item[c.] A voxel-based box-counting method was utilized to compute the fractal (box-counting) dimensions of the generated structures. The estimated non-integer dimensions confirmed the presence of genuine fractal geometry in the BLP solutions.

    \item[d.] Dimension values decreased slightly with successive spatial magnification, indicating variation in local complexity and revealing how structural features evolve under scale refinement.

    \item[e.] The study highlights the rich geometrical behavior of the BLP system by integrating analytical solution construction with fractal dimension theory, offering new insights into the interplay between nonlinearity, dispersion, and pattern formation.
\end{enumerate}

In conclusion, this work affirms the capacity of the BLP system to generate fractal waveforms governed by nonlinear dispersive dynamics. The combination of visual and metric-based approaches provides a robust framework for understanding self-similar structures in nonlinear PDEs. These findings not only enhance the theoretical understanding of the BLP system but also suggest potential applications in modeling fine-scale patterns in fluid dynamics, nonlinear optics, and wave turbulence. Future research may extend these results by exploring multifractal properties, temporal evolution of fractal structures, or applying the methodology to other integrable and non-integrable dispersive systems.

\section{Acknowledgment:} 
Financial support provided by the University Grants Commission (UGC), India, is gratefully acknowledged by Saugata Dutta (NTA Ref. No. $211610066362$) for enabling to fulfill  this work.  


 

\end{document}